\documentclass[aps,prl,showpacs,citeautoscript,twocolumn]{revtex4}

\usepackage{graphicx}
\usepackage{amsmath}
\usepackage{amssymb}

\newcommand{\bnen}{\begin{equation}}
\newcommand{\eden}{\end{equation}}
\newcommand{\bean}{\begin{eqnarray}}
\newcommand{\eean}{\end{eqnarray}}
\newcommand{\spinor}[2]{\left(\bna{c} #1 \\[1.5ex] #2 \eda\right)}
\newcommand{\bna}{\begin{array}}
\newcommand{\eda}{\end{array}}
\newcommand{\J}{\mathbb{J}}
\newcommand{\f}{\frac}

\newcommand{\phii}{\varphi_{i}}

\newcommand{\sigmavec}{\mbox{\boldmath{$\sigma$}}}

\begin{document}
\title{Caustics due to Negative Refractive Index in Circular 
Graphene {\em p-n} Junctions}

\author{J\'ozsef Cserti}

\author{Andr\'as P\'alyi}

\author{Csaba P\'eterfalvi}
\affiliation{
E\"otv\"os University, Budapest, Hungary, 
\\ H-1117 Budapest, P\'azm\'any P{\'e}ter s{\'e}t\'any 1/A, Hungary}


\begin{abstract}
We show that the wavefunctions form caustics in circular graphene
{\em p-n} junctions which in the framework of geometrical optics 
can be interpreted with negative refractive index.

\end{abstract}


\maketitle


The possibility of negative refractive index in nature was 
first analyzed theoretically by Veselago\cite{Veselago:cikk} which was
followed by other important works\cite{Pendry1:cikk_Pendry2:cikk_Smith:cikk}. 
Evidence for such left-handed metamaterials has been demonstrated 
by microwave experiments\cite{Shelby:cikk_Houck:cikk_Parazzoli:cikk_Notomi:cikk_Luo:cikk}. 

Another prominent candidate for such materials might be the graphene
as it was proposed recently by Cheianov et al. ~\cite{Falko_optics:cikk}. 
They studied the transmission of electrons through a plane {\em p-n}
junction of graphene and showed that the optics of electron flow, 
in the framework of geometrical optics, can be described by optical 
refraction with negative refractive index. 
The graphene itself provides many other peculiar electronic properties owing
to the close similarity between the dispersion relation  
of two-dimensional massless Dirac fermions and the low energy
electronic spectrum of graphene (for review 
see~\cite{review_graphene_cikkek}). 
However, an easy way of tuning the refractive index by gate potential  
may open up further research directions in graphene physics. 
For example, graphene might be utilized to fabricate properly designed 
electronic lens. 

One of the interesting subjects in geometrical optics is the caustics.  
A caustic is an envelope of a family of rays at which 
the density of rays is singular.
The caustics have been extensively studied
in the past (for review see, eg, the work by Berry and
Upstill~\cite{Berry_Upstill:cikk}). 
As in Ref.~\cite{Falko_optics:cikk} for plane {\em p-n} junction, 
we show theoretically that in circular {\em p-n} junction of
graphene the caustics can also arise in the wavefunction pattern of 
electrons, and the curves of caustics can be calculated 
using the well-known Snell's law with negative refractive index.  

\begin{figure}[hbt]
\includegraphics[scale=0.4]{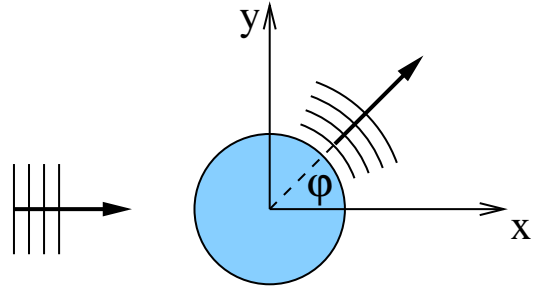}
\caption{\label{setup:fig}  
Incident plane wave of electron in single layer graphene is scattered 
by a rotational symmetric potential $V(r)$ (circular region).
}
\end{figure}

To this end, we consider the scattering of ballistic incident electrons in
graphene shown in Fig.~\ref{setup:fig}.
To demonstrate the formation of caustics in circular {\em p-n}
junctions, we used a simple gate potential 
$V(r) = V_0 \Theta(R-r)$, where $\Theta$ denotes the Heaviside
function. 
The assumption of such a sharp potential is valid provided that
$\lambda_F \gg d$, where $\lambda_F$ is the Fermi wavelength outside
the junction and $d$ is the characteristic length scale
in which the scattering potential varies. 
Moreover, to prevent intervalley scattering it is neccessary that 
$d \gg a$, where  $a$ is the lattice constant of 
graphene~\cite{Falko_optics:cikk,Katsnelson-Klein:cikk}.
The negative refractive index is realized by {\em p-n} junctions of
graphene in which the energy $E$ of the incoming electrons 
at the {\em n} side is positive 
(belonging to the conduction band), while at the {\em p} side of the
junction the potential $V_0$ is chosen as $E<V_0$, therefore the
electrons belong to the valence band. 
Here we focus on this case ($0<E<V_0$) but its generalization 
to arbitrary values of $E$ is straightforward.
Note that various properties of elastic electron scattering 
in graphene has already been studied~\cite{szoras_graphene:cikkek}. 
Here we investigate the wavefunction pattern inside the junction. 
Pronounced formation of the caustics can only be seen 
in the quasi-classical limit, ie, in case of $R\gg \lambda_F$, 
but no such condition is assumed in our exact calculation of 
the wavefunctions.      

The Hamiltonian for the above described scattering problem for energies
$E$ close to the Dirac point may be written as 
\bnen
H = H_0 + V({\bf r})  \openone 
= c \ \sigmavec \cdot {\bf p} + V({\bf r}) \openone .  
\label{Hamiltonian:eq}
\eden
Here $c$ is the Fermi velocity, 
${\bf p}= -i \hbar \,\partial/\partial {\bf r}$, while 
$\sigmavec = (\sigma_x,\sigma_y)$ and $\openone $ are 
the Pauli matrices and the unit matrix acting in isospin space. 

Consider the elastic scattering of incoming electrons governed by the
Hamiltonian $H$. 
The scattering of the incident electron can be calculated by 
the generalization of the well-known partial wave method 
(see eg, \cite{Schwabl:book_adhikari:cikk}). 
The wavefunction (in polar coordinates) describing 
the scattering of a single incoming partial wave $h^{(2)}_j$ (defined below)  
outside the scattering region ($r>R$) can be written in terms of 
the cylindrical wave eigenfunctions of the Hamiltonian $H$ 
with energy $E>0$:
\begin{subequations}
\begin{eqnarray}
\label{eq:cylindricalwaves}
\Psi^{\rm (out)}_j &=& h_j^{(2)} + S_j h_j^{(1)},\,\,\, \text{where}  \\
h_{j}^{(d)} (r,\varphi) &=&  
\spinor{H_{j-\frac{1}{2}}^{(d)}(k_{\text{out}}r)e^{-i\varphi/2}}
{i H^{(d)}_{j+\frac{1}{2}}(k_{\text{out}}r) 
e^{i\varphi/2}} e^{ij\varphi}, 
\end{eqnarray}%
\end{subequations}%
while inside the scattering region ($r<R$) the wavefunction reads 
\begin{subequations}
\begin{eqnarray}
\Psi^{\rm (in)}_j &=& A_j \chi_j , \,\,\, \text{where} \\
\chi_j(r,\varphi) &=& 
\spinor{J_{j-\frac{1}{2}}(k_{\text{in}}r)e^{-i\varphi/2}}{-iJ_{j+\frac{1}{2}}(k_{\text{in}}r)e^{i\varphi/2}}
e^{ij\varphi}. 
\end{eqnarray}%
\end{subequations}%
Here the pseudo angular momentum 
$j \in \J \equiv \{\dots,-\f32, -\f12, \f12, \f32,\dots\}$,
$h_{j}^{(1)}$ ($h_{j}^{(2)}$) is an outgoing (incoming) 
cylindrical wavefunction corresponding to $d=1$ ($d=2$). 
The wavenumbers are $k_{\text{out}}=E/(\hbar c)> 0$ 
and $k_{\text{in}}=|E-V_0|/(\hbar c) > 0$, 
while $J_n$ refers to the Bessel function of the first kind, and 
$H_n^{(1)}$ and $H_n^{(2)}$ are the Hankel functions of the first 
and second kind, respectively\cite{abramowitz:book}.
To construct the eigenfunctions we used the fact that 
for rotational symmetric potential $[J_z,H] = 0$ holds, 
where $J_z = -i\hbar \partial_\varphi + \hbar \sigma_z / 2$ is the 
pseudo angular momentum operator.
Therefore, the pseudo angular momentum is a conserved
quantity in the scattering process. 

The scattering matrix  $S_j$  and the amplitude $A_j$ are determined
from the boundary conditions, ie, from the continuity of
the total wavefunction at the boundary of the junction: 
$\Psi^{\rm (out)}_j(r=R,\varphi) = \Psi^{\rm (in)}_j(r=R,\varphi)$.
Then, it is easy to find that 
\begin{subequations}
\bean
S_j \!\! &=& \!\!\! 
\frac{-H_{j-\frac{1}{2}}^{(2)}(X) J_{j+\frac{1}{2}}(X^\prime)
- H_{j+\frac{1}{2}}^{(2)}(X) J_{j-\frac{1}{2}}(X^\prime) }{D_j}, \\[1ex]
A_j \!\! &=& \!\! \frac{H_{j-\frac{1}{2}}^{(2)}(X)  H_{j+\frac{1}{2}}^{(1)}(X)
- H_{j+\frac{1}{2}}^{(2)}(X) H_{j-\frac{1}{2}}^{(1)}(X) }{D_j}, \\[1ex]
D_j \!\!\! &=& \!\!\!  H_{j+\frac{1}{2}}^{(1)}(X) J_{j-\frac{1}{2}}(X^\prime)
+ H_{j-\frac{1}{2}}^{(1)}(X) J_{j+\frac{1}{2}}(X^\prime), 
\eean%
\label{S_A:eq}%
\end{subequations}%
where $X= k_{\text{out}}R$ and $X^\prime = k_{\text{in}}R$. 

We now consider the scattering of an incident plane wave of electron
in graphene for $r>R$. 
Such an eigenstate with energy $E$ has the form
\begin{subequations}
\bean
\label{eq:planewave}
\Phi_{\phii}(r,\varphi) &=&  \eta(\phii) 
e^{ik_{\text{out}}r\cos(\varphi-\phii)},
\,\,\, \text{where} \\
\eta(\phii) &=& \frac{1}{\sqrt{2}}\,\spinor{e^{-i\phii/2}}{e^{i\phii/2}}.
\eean%
\end{subequations}%
and $\phii$ denotes the direction of the propagation of 
the incident electron.  
Using the properties of the Hankel functions~\cite{abramowitz:book} one
can show that the partial wave expansion of $\Phi_{\phii}$ is 
\bnen
\label{eq:partial}
\Phi_{\phii} = \f 1 2 \sum\limits_{j\in\J} i^{j-\frac{1}{2}}
 (h_j^{(2)} + h_j^{(1)}) e^{-ij\phii}.
\eden
Without the loss of generalization, we choose the direction of propagation to 
be parallel with the $x$ axis, which means $\phii = 0$ in
\eqref{eq:planewave}.
Then the wavefunction for $r>R$ is given by 
\begin{subequations}
\label{eq:psiinout}
\bnen
\label{eq:psiout}
\Psi^{\rm (out)} = \Phi_{\phii =0} + 
\f 1 2 \sum\limits_{j\in\J}  i^{j-\frac{1}{2}} (S_j -1)
h_{j}^{(1)}, 
\eden
and the wavefunction for $r<R$ is given by
\bnen
\label{eq:psiin}
\Psi^{\rm (in)} = \f 1 2 \sum\limits_{j\in\J} i^{j-\frac{1}{2}} A_j \chi_j.
\eden%
\end{subequations}%
Note that the second term in (\ref{eq:psiout}) is the scattered wave
due to the scattering of the incident plane wave $\Phi_{\phii}$ on
the scattering region described by the potential 
$V(r) = V_0 \Theta(R-r)$. 
The scattering cross section can be obtained from the asymptotic 
form ($r\gg R$) of the scattered wave. 

\begin{figure}[hbt]
\includegraphics[scale=2.2]{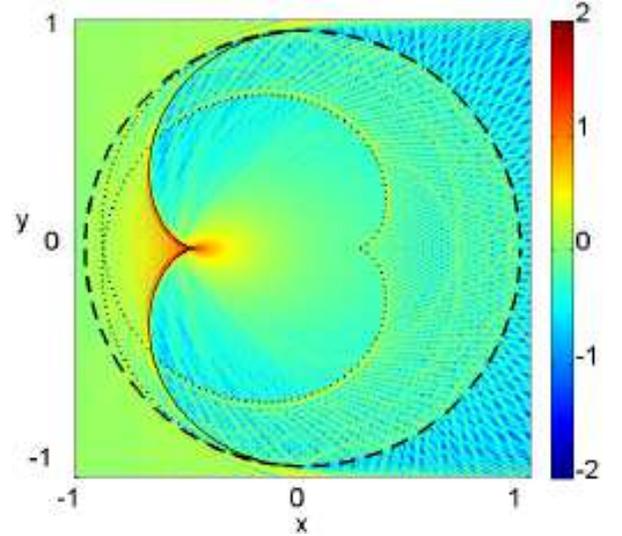}
\caption{\label{wavefn-1:fig}  
$|\Psi|^2$ (in scale of logarithmic to base $10$) is plotted 
inside and outside, close to the scattering
region (the dashed line shows the boundary of the {\em p-n} junction). 
Here $k_{\text{in}}R= 300$ and $k_{\text{out}}R= 300$ 
corresponding to $n= -1$, and $x$ and $y$ are in units of $R$. 
The solid (dotted) line corresponds to the caustic for $p=1$ ($p=2$)
(see the text). 
}
\end{figure}

\begin{figure}[hbt]
\includegraphics[scale=2.2]{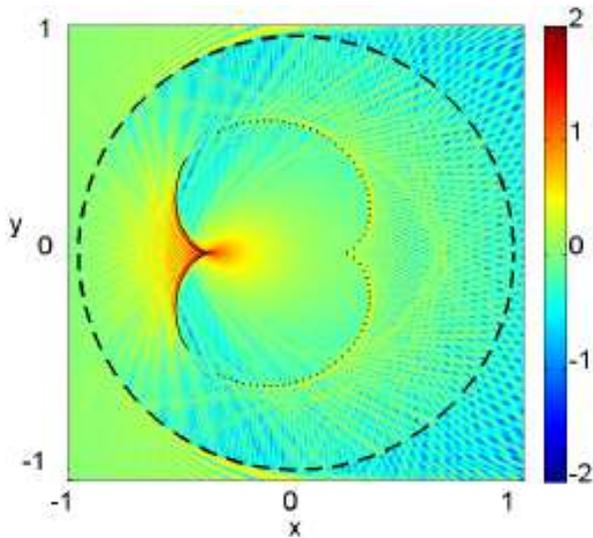}
\caption{\label{wavefn-2:fig} 
The same as in Fig.~\ref{wavefn-1:fig} with 
$k_{\text{in}}R= 300$ and $k_{\text{out}}R= 200$ corresponding to 
$n= -1.5$. 
}
\end{figure}

Equations~(\ref{S_A:eq}) and (\ref{eq:psiinout}) allows us to
calculate exactly the wavefunctions both inside and outside the
junction. 
Note that the wavefunctions depend only on the two dimensionless
parameters, $k_{\text{in}}R$ and $k_{\text{out}}R$. 
Figures~\ref{wavefn-1:fig} and \ref{wavefn-2:fig} show how the
incident plane wave (from direction $\varphi_i = 0$) penetrates into
the circular region of the {\em p-n} junction. 
For $r<R$, the formation of the caustics in the wavefunction patterns
is clearly visible along the solid and dotted lines.   
Vodo et al. found a similar wave pattern experimentally by focusing a plane 
microwave with a plano-concave lens fabricated from a photonic crystal having
negative refractive index~\cite{plano-concave:cikk} . 

In a sharp {\em p-n} junction of graphene the optics of electron 
flow~\cite{Falko_optics:cikk} 
is very much the same as in photonic crystals with negative refractive index. 
In a similar way as in Ref.~\cite{Falko_optics:cikk} 
for our circular {\em p-n} junction the Snell's law reads 
\begin{equation}
\frac{\sin \alpha}{\sin \beta} = n = - \frac{k_{\text{in}}}{k_{\text{out}}},
\label{Snell:eq}
\end{equation}
where $\alpha$ and $\beta$ are the angle of incidence and refraction,
respectively. 
Since in our calculation both $k_{\text{in}}$ and $k_{\text{out}}$ 
are positive the refractive index becomes negative 
similarly as for properly designed photonic crystals.  

\begin{figure}[hbt]
\includegraphics[scale=0.45]{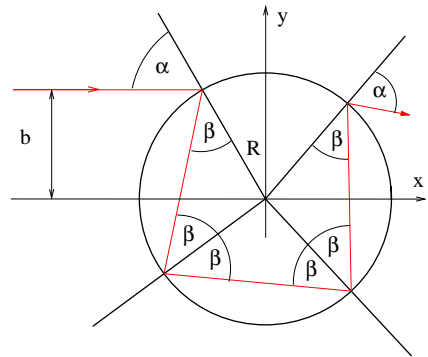}
\caption{\label{ray_expl:fig} 
The basic geometry of ray paths inside the circular {\em p-n}
junction. 
The incident ray from left with impact parameter $b$ and angle of
incidence $\alpha$ is refracted at the boundary of the junction with angle
$\beta$ and then after $p-1$ internal reflections exits from the junction. 
Here $p=3$ and we set $n=-1.3$ and $\alpha = 60^\circ$. 
}
\end{figure}

\begin{figure}[hbt]
\includegraphics[scale=0.7]{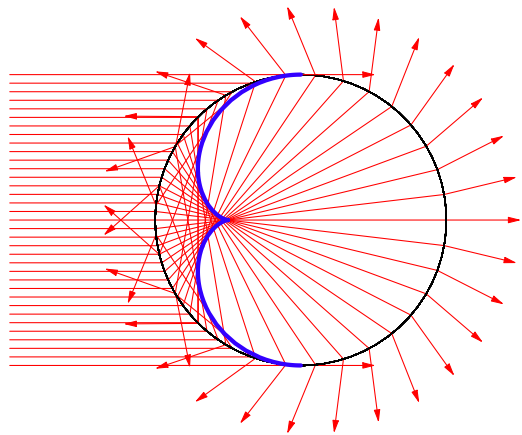}
\caption{\label{caustics_geo:fig} 
Caustic inside the {\em p-n} junction formed from the envelope of 
the refracted ray paths of different incident rays.
The thick solid line is the caustic curve calculated 
from (\protect \ref{caustics:eq}). 
Here $p=1$ and $n=-1$.
}
\end{figure}

In what follows we show that the intensity maximum in the wavefunction
patterns is around the caustic and can be understood by the Snell's
law (\ref{Snell:eq}) with negative refractive index.  
Figure~\ref{ray_expl:fig} shows how the incident ray refracts at
the boundary of the  circular {\em p-n} junction and then after $p-1$ 
internal reflections exits from the junction. 
One can classify the different ray paths by the impact parameter
$b=R\sin \alpha$ and the number of chords $p$
inside the circle corresponding to $p-1$ internal reflections.  
Incident rays with varying impact parameters ($-R \le b \le R$)
form a family of rays inside the circle. 
The envelope of this ray family results in a caustic as shown in
Fig.~\ref{caustics_geo:fig}. 
Each of the chords has its own caustic. 

Using differential geometry to calculate the envelope of family of
curves we find that the curve ${\bf r}_c$ (in Descartes coordinates
shown in Fig.~\ref{ray_expl:fig}) of the caustic of the $p$th chord 
is given by
\begin{subequations}
\begin{eqnarray}
\frac{{\bf r}_c(p,\alpha)}{R} &=& {(-1)}^{p-1} \Biggl[
\left(\bna{c} -\cos\Theta \\ \sin\Theta \eda\right)
\Biggr. \nonumber \\[2ex] 
&& \Biggl. \hspace{-10mm} + \cos \beta\, 
\frac{1+2\left(p-1\right)\beta^\prime}{1+\left(2p-1\right)\beta^\prime}\,
\left(\bna{c} \cos (\Theta+\beta) \\ -\sin (\Theta+\beta) \eda\right)
\Biggr] , \\[2ex]
\Theta(p,\alpha)  &=& \alpha + 2\left(p-1\right)\beta , \\ 
\sin \beta  &=& \frac{\sin \alpha}{|n|}, \\
\beta^\prime  &=& \frac{\cos \alpha}{\sqrt{n^2-\sin^2 \alpha}}. 
\end{eqnarray}%
\label{caustics:eq}%
\end{subequations}%
Here $\alpha$ varies between $-\pi/2$ and  $\pi/2$ and the prime
denotes the derivation with respect to $\alpha$.
This is a parametric curve for caustics with parameter $\alpha$.
Figures~\ref{wavefn-1:fig} and \ref{wavefn-2:fig} show the caustics  
calculated from Eq.~(\ref{caustics:eq}). 
In both cases, one can see that the location of the caustics formed
from the interference pattern of the exact wavefunctions inside 
the junction agrees very well with that obtained 
from Snell's law with negative refractive index.    
The caustics for $p>2$ are less visible since after each internal
reflection the intensity of the rays decreases. 

The caustics in a circular {\em p-n} junction belong to the class of cusp
according to the catastrophe optics~\cite{Berry_Upstill:cikk}. 
The location of the cusp $(r,\varphi)=(r_{\text{cusp}},\pi)$ 
(in polar coordinates) for the $p$th chord 
can be obtained from (\ref{caustics:eq}) 
by setting $\alpha =0$ and we find  
\begin{equation} 
r_{\text{cusp}}(p) = \frac{{(-1)}^{p} }{|n|-1 +2p}\, R. 
\label{cusp:eq}
\end{equation} 
As can be seen in Fig.~\ref{caustics_geo:fig} the paraxial 
($\alpha \ll 1$) incident rays entering into the circular 
region of the graphene junction are focused at the focal point for $p=1$. 
The focal length $f$ measured from the point $(r,\varphi)=(R,\pi)$ 
is given by the location
of the cusp: $f = R -|r_{\text{cusp}}|=  R |n|/(|n|+1)$.
The same is obtained when the refractive index is replaced by its negative
value in the expression of the image focal length defined in ordinary
geometrical optics~\cite{Hecht:book}.  
To demonstrate the negative refractive index in the photonic crystal
experiment\cite{plano-concave:cikk} the focal length was measured. 

To realize the predicted caustic formation and focusing effect 
in an experiment, one needs to collimate a monodirectional electron
beam onto the circular scattering region. 
This might be achieved utilizing a \emph{smooth planar p-n} junction  
which is known to transmit only those quasiparticles that approach 
it almost perpendicular to the {\em p-n} 
interface~\cite{Cheianov_Falko_selective:cikk}. 
Therefore, a possible experimental setup could be built up 
from a source electrode, a selectively transmitting smooth planar 
{\em p-n} junction, the circular scattering region and a drain electrode. 
Under bias, the spatial dependence of the charge density of the transported 
electrons might be measured by scanning probe techniques.

Circular {\em p-n} junctions together with planar ones 
studied earlier~\cite{Falko_optics:cikk} can be a building 
block of electron optics in graphene. 
However, the refractive index varies with energy of the incoming
electrons therefore the temperature needs to be low enough for sharp images.
As it is mentioned in the introduction, one needs experimentally that 
$a \ll d \ll \lambda_F \equiv 2\pi/k_{\text{out}}$ 
(for sharp potential barrier and absence of intervalley scattering). 
Another condition required to observe sharp interference patterns
around the caustics is $k_{\text{in}} R \gg 1$ corresponding to the 
quasi-classical limit. 
For example, setting $E=40$~meV, and $V_0=80$~meV (using gate potential), 
$R = 800$~nm, all of these conditions are fullfiled since $d\sim
10$~nm\cite{Huard_exp_barrier:cikk}, 
implying  that the refractive index is $n=-1$, while 
$k_{\text{in}} R = 50$, and $\lambda_F = 10d$.

In summary, we calculated inside and outside a circular {\em p-n} 
junction of graphene the scattered wavefunction of an incoming 
plane wave of electrons due to a circular symmetric step-like potential. 
We showed that the scattered wavefunction inside the junction 
has an interference pattern with high intensity maximum located
around the caustics calculated from Snell's law with negative refractive
index.

We gratefully acknowledge discussions with 
B. L. Altshuler, C. W. J. Beenakker, V. V. Cheianov, V. Fal'ko, 
F. Guinea, and A. Zawadowski.
This work is supported by European Commission Contract No.~MRTN-CT-2003-504574.



\end{document}